\documentclass[12pt]{iopart}
\usepackage{epsfig}
\usepackage{iopams}
\usepackage{graphicx}% Include figure files
\usepackage{dcolumn}% Align table columns on decimal point
\usepackage{bm}% bold math
\usepackage{color}
\begin{document}

\title[FIR optical conductivity of CeCu$_{2}$Si$_{2}$]{Far-infrared optical conductivity of CeCu$_2$Si$_2$}

\author{J. Sichelschmidt, A. Herzog, H.S. Jeevan, C. Geibel, F. Steglich}
% \altaffiliation[Electronic address: ]{Sichelschmidt@cpfs.mpg.de}
%\author{A. Herzog}
%\author{H. Jeevan}
%\author{C. Geibel}
%\author{F. Steglich}
%\affiliation{Max Planck Institute for Chemical Physics of Solids, N\"othnitzer Stra\ss e 40, 01187 Dresden, Germany}
\address{Max Planck Institute for Chemical Physics of Solids, N\"othnitzer Stra\ss e 40, 01187 Dresden, Germany}
\ead{Sichelschmidt@cpfs.mpg.de}
\author{T. Iizuka, S. Kimura}
%\affiliation{UVSOR Facility, Institute for Molecular Science, Okazaki 444-8585, Japan}
%\affiliation{School of Physical Sciences, The Graduate University for Advanced Studies (SOKENDAI), Okazaki 444-8585, Japan}
\address{
UVSOR Facility, Institute for Molecular Science, Okazaki 444-8585, Japan
}
\address{
School of Physical Sciences, The Graduate University for Advanced Studies (SOKENDAI), Okazaki 444-8585, Japan
}

\date{DRAFT \today}

\begin{abstract} 

[Journal ref.: J. Phys.: Condens. Matter {\bf 25}, 065602 (2013)] We investigated the optical reflectivity of the heavy-fermion metal CeCu$_2$Si$_2$ in the energy range 3 meV -- 30 eV for temperatures between 4K -- 300K. The results for the charge dynamics indicate a behavior that is expected for the formation of a coherent heavy quasiparticle state: Upon cooling the spectra of the optical conductivity indicate a narrowing of the coherent response. Below temperatures of 30~K a considerable suppression of conductivity evolves below a peak structure at 13~meV. We assign this gap-like feature to strong electron correlations due to the 4$f$-conduction electron hybridization.
\end{abstract}

\pacs{71.27.+a, 78.20.-e}

%%%%%%%%%%%%%%%%%%%%%%%%%%%%%%%%%%%%%%%%%%%%%%%%%%%%%%%%%%%%
\maketitle
\section{Introduction}
The investigation of the optical properties of heavy fermion metals with $4f$-electrons is focussed on the charge dynamics in the far- and mid-infrared energy region where the electronic structure is strongly influenced by a weak hybridization of localized $f$-electrons with conduction electrons. The relevant energy scales involved are determined by the Kondo interaction, leading to a strongly enhanced density of states at the Fermi level and the concomitant formation of heavy quasiparticle masses. A heavy fermion state arises upon the formation of coherence among heavy quasiparticles below a lattice coherence temperature. The latter has been shown to be characteristic for the temperature dependence of the far-infrared optical response in heavy fermion compounds like YbRh$_{2}$Si$_{2}$ and YbIr$_{2}$Si$_{2}$ \cite{kimura06a,sichelschmidt08a,iizuka10a}. The signature of the Kondo interaction itself was discussed in terms of a gap-like suppression in the far-infrared optical conductivity due to a conductionband ($c$)-$f$ electron hybridization. This suppression appears at temperatures below the Kondo temperature and is related to heavy fermion properties determining, for instance, the thermodynamics. Furthermore, the strength of the $c\,$-$f$ hybridization is reflected in a broad mid-infrared peak in the range $0.1-0.3$~eV at low temperatures \cite{okamura07a}. It was shown that a dynamical mean-field approach could capture these $c\,$-$f$ hybridization features in the far-infrared region (where several maxima depending on the complexity of the hybridization could appear) as well as in the mid-infrared region\cite{shim07a,kumar11a}.\\
Although CeCu$_{2}$Si$_{2}$ is one of the most investigated heavy-fermion compounds \cite{stockert12a} yet, to the best of our knowledge, it does not belong to the large number of correlated electron materials whose optical properties were reported \cite{basov11a}. Here we report optical investigations of CeCu$_{2}$Si$_{2}$ down to energies of 3~meV and temperatures down to 4~K.
This allowed us to obtain yet inaccessible information on the low-energy heavy-fermion optical response of CeCu$_{2}$Si$_{2}$.

%------------------------------------------------------------------------------
%
\section{Experimental}
\indent We have measured the optical reflectivity $R(\omega)$ at a near-normal angle of incidence on a single crystal of CeCu$_{2}$Si$_{2}$. The crystal was well characterized in its thermodynamic, transport and magnetic properties and shows superconductivity below $T_{\rm C}=0.65$~K as well as antiferromagnetic order below $T_{\rm N}=0.6$~K (''S/A-type'' crystal) \cite{jeevan10a}. Those properties below 1~K do not affect the transport properties at higher temperatures, hence the DC (zero energy) electrical resistivity $\rho_{\rm DC}$ shows the typical temperature dependence for all types of CeCu$_{2}$Si$_{2}$ samples, namely the presence of two maxima, one at 20~K due to the onset of quasiparticle coherence and one at 115~K being due to the Kondo scattering at a quasi-quartet crystalline electric field level \cite{jeevan10a}.
The direction of the crystalline $c$-axis (tetragonal ThCr$_{2}$Si$_{2}$ structure) was pointing by an angle of $15\pm6^{\circ}$ out of the optically investigated crystal surface. The well polished ($0.3\,\mu \rm m$ grain) crystal was mounted behind a circular aperture with a diameter of 5~mm.\\
A rapid-scan Fourier spectrometer of Michelson and Martin-Puplett type was used for photon energies $\hbar\omega$
between of 0.01--1.5~eV and 3--30~meV, respectively, at sample temperatures between 4--300~K. $R(\omega)$ was extracted from two measurements: the first one measuring the sample and the second one measuring the sample after coating it \textit{in-situ} with gold. Synchrotron radiation extended the energy range from 1.2\,eV up to 30\,eV for $T=300$~K \cite{fukui01a}. Using Kramers-Kronig relations, we calculated the optical conductivity $\sigma(\omega)$ from $R(\omega)$. The $R(\omega)$ spectra were extrapolated below 3~meV with the low-energy limit of the Drude model $R(\omega)=1-(2\omega\rho_{\rm DC}/\pi)^{1/2}$ (Hagen-Rubens behavior) and above 30~eV with a free-electron approximation $R(\omega) \propto \omega^{-4}$ \cite{dressel02a}.

%%%%%%%%%%%%%%%%%  FIG.1  %%%%%%%%%%%%%%%%%%%%%%%%
\begin{figure}[hbt]
\begin{center}
\includegraphics[width=0.7\columnwidth]{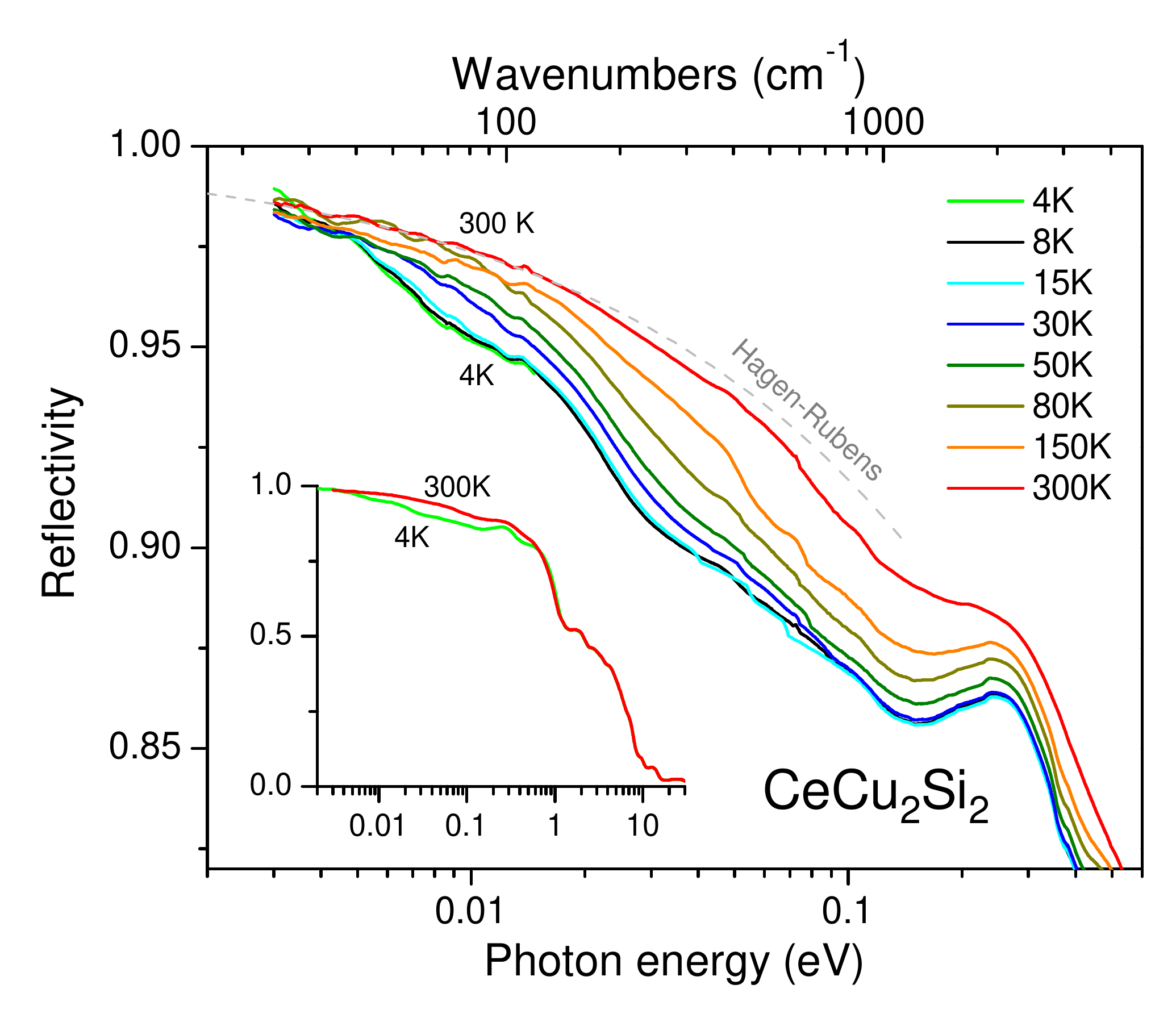}
\end{center}
\caption{
(Color online) Temperature dependence of the reflectivity spectrum $R(\omega)$. Dashed line shows Hagen-Rubens behavior according to a DC resistivity of $70\, \mu\Omega \rm{cm}$.
Inset: $R(\omega)$ at 4 and 300~K in the complete accessible range of photon energies up to 30~eV.
}
\label{fig1}
\end{figure}
%%%%%%%%%%%%%%%%%%%%%%%%%%%%%%%%%%%%%%%%%%%%%

%------------------------------------------------------------------------------
%
\section{Results and Discussion}
%
%%%%%%%%%%%%%%%%  FIG.2n  %%%%%%%%%%%%%%
\begin{figure}[b]
\begin{center}
\includegraphics[width=0.7\columnwidth]{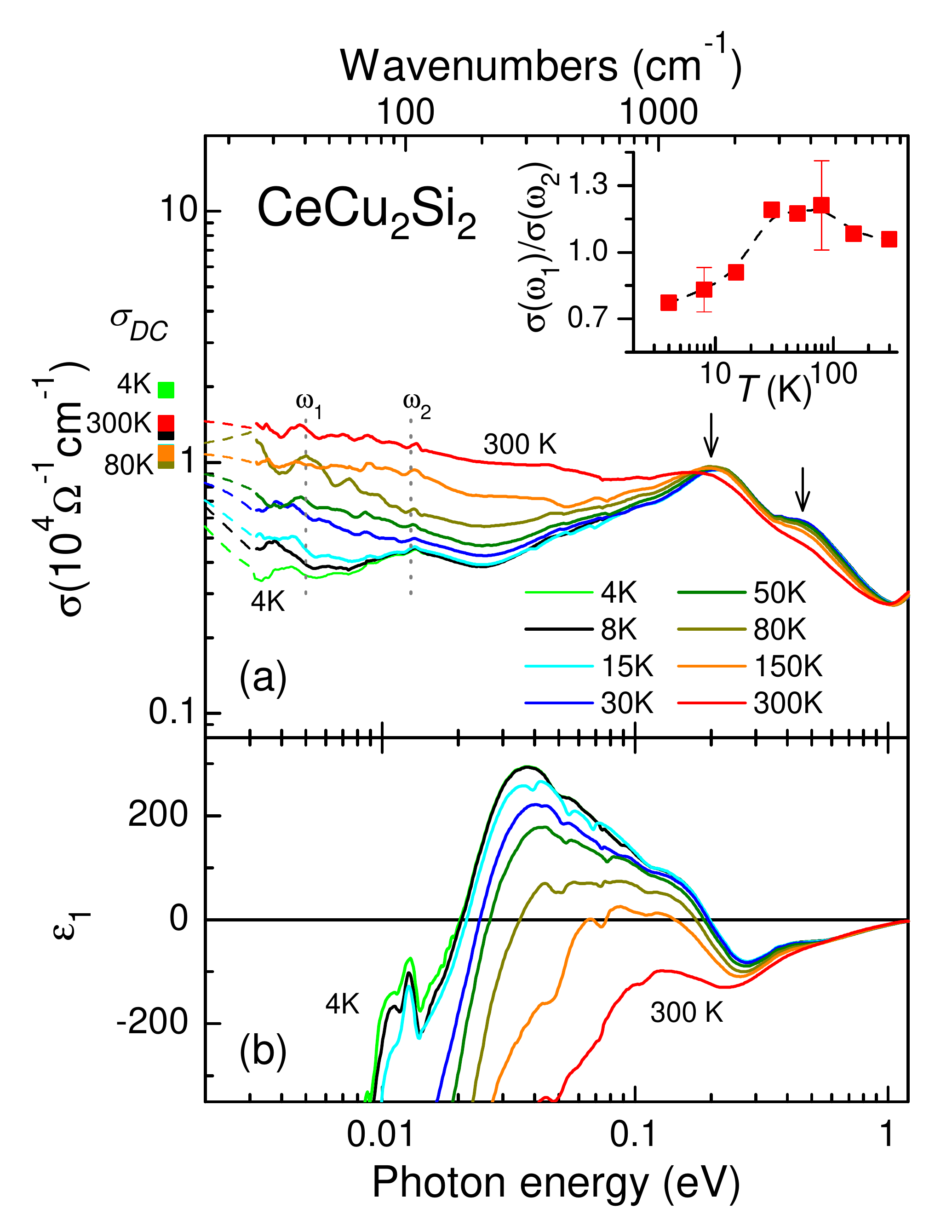}
\end{center}
\caption{
(Color online) (a) Temperature dependence of the optical conductivity $\sigma(\omega)$ with corresponding direct current conductivity (symbols: $\sigma_{DC}$ - values at zero energy). Arrows: typical signatures related to the $c\,$-$f$ hybridization strength in Ce- and Yb- heavy-fermion compounds.
Inset: Temperature dependence of $\sigma(\omega)$ at 5~meV normalized to the 13~meV values (dashed line guides the eyes; error bars, being representative for different temperature ranges, result from experimental accuracies of $\sigma_1(\omega_{1,2})$).
(b) Real part of the complex dielectric function $\varepsilon_{1}(\omega)$ revealing the heavy plasmon energy from zero-crossings.}
\label{fig2}
\end{figure}
%%%%%%%%%%%%%%%%%

\indent The temperature dependence of the $R(\omega)$ spectra of CeCu$_{2}$Si$_{2}$ is shown in Fig.~\ref{fig1}. 
The continuous increase of $R(\omega)$ towards low energies indicates the response of the conduction band. The dashed line shows a Hagen-Rubens behavior with $\rho_{\rm DC}^{300K}=70\, \mu\Omega \rm{cm}$ (in good agreement with the measured absolute value of $\rho_{\rm DC}(T)$ \cite{jeevan10a,trovarelli97a}) providing a reasonably well description of the low-energy data. However, inspecting the temperature evolution of $R(\omega)$ typical metallic behavior is seen only for $\omega\lesssim5$~meV and below $T\lesssim30$~K. This peculiarity is often reported for heavy fermion metals and is discussed in terms of a renormalization of the coherent Drude response \cite{basov11a}.

Fig.~\ref{fig2}(a) shows the temperature dependence of the optical conductivity $\sigma(\omega)$ below 1~eV where the charge dynamics is characterized by a pronounced temperature dependence. Between the lowest accessible energy of 3~meV and 150~meV, $\sigma(\omega)$ shows a continuous decrease with decreasing temperature. As shown in Fig. \ref{fig2}(b), this energy range is characterized by the formation of a heavy plasma mode. For heavy-fermion compounds, in the low-temperature coherent state, the real part of the complex dielectric function, $\varepsilon_{1}(\omega)$, is expected to show two zero-crossings from negative to positive. The crossing at lower energy ($\omega^*_p$) is associated with heavy plasmons, the one at higher energy with the plasma energy of uncorrelated electrons \cite{millis87b}. Such analysis of $\varepsilon_{1}(\omega)$ was considered, for instance, for CeAl$_{3}$ \cite{awasthi93a} or for the Ce-115 system \cite{mena05a}.
For CeCu$_{2}$Si$_{2}$, see Fig. \ref{fig2}(b), the heavy plasma mode $\omega^*_p$ develops already at $\approx150$~K and reaches values down to about 20~meV for 4~K. Since the effective mass $m^*$ is proportional to 1/$\omega^{*2}_p$, $m^*$ at 4~K is about ten times larger than that at 150~K. The plasma energy for the uncorrelated electrons is located at $\approx1.5$~eV. 

Additional information on the heavy fermion optical response and its relation to the Kondo interaction can be inferred from the behavior of $\sigma(\omega)$ at the lowest energies. Interestingly, for $T<30$~K $\sigma(\omega)$ shows an additional suppression in the region around $\omega_{1}=5$~meV and a concomitant evolution of a peak structure at $\omega_{2}\approx13$~meV, i.e. a gap-like feature appears. As illustrated in the inset this feature shows a temperature dependence which we estimated by using the temperature evolution of $\sigma(\omega_{1})/\sigma(\omega_{2})$. The overall temperature behavior of this ratio does not depend on the exact choice of $\omega_{1}$ and $\omega_{2}$. From 300~K down to 30~K $\sigma(\omega_{1})/\sigma(\omega_{2})$ shows a continuous increase reflecting a continuous decrease of the scattering rate of the low-energy Drude part. Below 30~K, towards lower temperatures, a strong decrease of $\sigma(\omega_{1})/\sigma(\omega_{2})$ demonstrates the gap opening resulting in an additional suppression of $\approx25\%$ for $\sigma(\omega_{1})$ at the lowest temperature. Note, that in this temperature regime also the Kondo effect with $T_{\rm K}=10$~K should influence the charge dynamics, suggesting a relation of this gap to the presence of 4$f$ electrons. Hence, we attribute the suppression of $\sigma(\omega)$ below 13~meV to the opening of a hybridization gap which arises from the weak hybridization of the 4$f$ states with the conduction electron band states as a consequence of the Kondo interaction. 

The properties of the peak structure at around 13~meV are similar to those observed in other 4$f$-based correlated electron systems, notably in the Kondo-semiconductors YbB$_{12}$ \cite{okamura05a} or CeFe$_{2}$Al$_{10}$ \cite{kimura11b}, the intermediate valent system YbAl$_{3}$ \cite{okamura04a}, and in the heavy-fermion systems YbRh$_{2}$Si$_{2}$ and YbIr$_{2}$Si$_{2}$ \cite{sichelschmidt08a}. For the skutterudite heavy-fermion system CeRu$_{4}$Sb$_{12}$ a peak of similar type is not reported with comparable clarity \cite{dordevic01a} whereas in certain Ce-based Kondo semiconductors comparable structures where discussed in terms of a pseudogap opening in the heavy quasiparticle band \cite{matsunami02a}. CeCu$_{2}$Si$_{2}$ seems to be the first example of a Ce-based heavy-fermion compound where such a clear gap opening in the vicinity of the Kondo temperature is observed. 

The peak and shoulder structures at around 0.2~eV and 0.45~eV (indicated by arrows in Fig. \ref{fig2}(a)) correspond to those typically found in a variety of Ce- and Yb- heavy fermion compounds. It was shown that their peak energies universally scale with the $c\,$-$f$ hybridization strength \cite{okamura07a}.

%
%%%%%%%%%%%%%%%  FIG.3  %%%%%%%%%%%%%%%
\begin{figure}[hbt]
\begin{center}
\includegraphics[width=0.7\columnwidth]{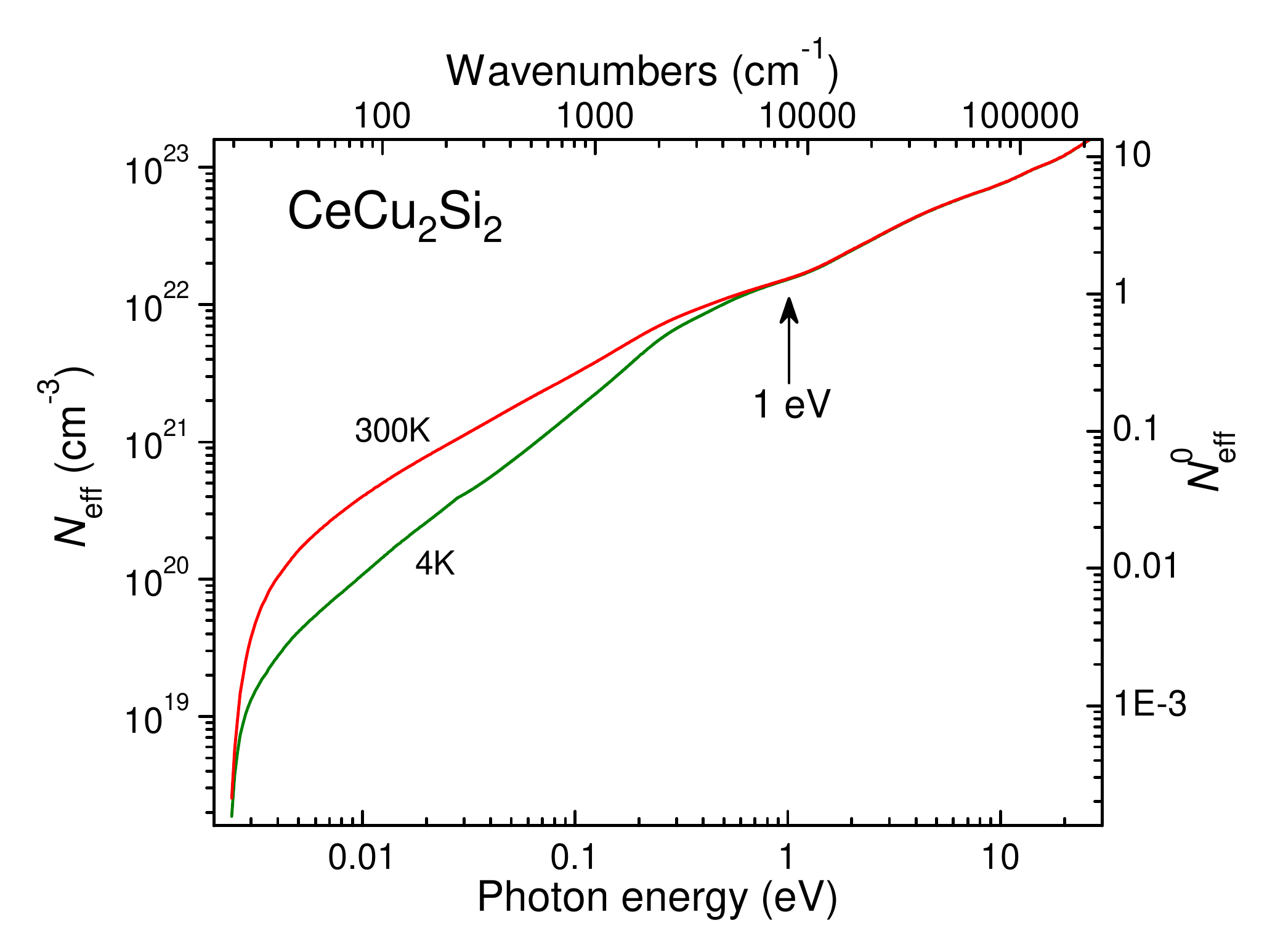}
\end{center}
\caption{
(Color online) Effective charge carrier density $N_{\rm eff}$ (from Eq.\ref{SW}) which contributes to optical excitations at a given energy. Below energies of $\approx 1$~eV a pronounced loss of spectral weight is visible upon cooling from 300~K to 4~K. Right axis shows $N_{\rm eff}^{0}=N_{\rm eff}\cdot V_{0}/2$ per formula unit using the unit cell volume $V_{0}=1.6695\cdot 10^{-22} {\rm cm}^3$.
}
\label{fig4}
\end{figure}
%%%%%%%%%%%%%%%%%%%%%%%%%%%%%%

The discussed signatures of strong electron correlations can be supplemented by a consideration of the energy dependence of the spectral weight which is determined by the effective charge carrier density $N_{\rm eff}$:
\begin{equation}
N_{\rm eff}(\omega)=\frac{2m_{0}}{\pi e^{2}} \int_{0}^{\omega}\sigma_{1}(\omega')d\omega'
\label{SW}
\end{equation}
with $m_{0}$ the free-electron mass. As shown in Fig.~\ref{fig4} $N_{\rm eff}$ becomes temperature independent for energies larger than 1~eV. This points to a correlation effect \cite{rozenberg96a} because it indicates a redistribution of spectral weight up to energies much larger than the hybridization gap energy of 13~meV which characterizes the low-temperature optical response of the heavy quasiparticles.
\section{Conclusion}
\indent The optical conductivity of the heavy-fermion metal CeCu$_{2}$Si$_{2}$ shows features typical for a heavy electron optical response. At the lowest accessed temperature of $T=4$~K and the lowest energy of 3~meV signatures of the tail of a renormalized Drude peak could be observed. This is suggested by a energy dependent suppression of the optical conductivity with decreasing temperature, indicating the effect of strong electron correlations. We identified a correlation induced hybridization gap which forms below about 13~meV and for temperatures below about 30~K. This agrees with the characteristic energy of the 4$f$-conduction electron interaction in CeCu$_{2}$Si$_{2}$ with a Kondo temperature of 10~K. The strongly temperature dependent gap formation in the optical conductivity leads to a transfer of spectral weight toward energies up to 1~eV.

%%%%%%%%%%%%%%%%%%%%%%%%%%%%%%

\section*{Acknowledgements}
We acknowledge S. Seiro for her assistance with the Laue X-ray diffraction of the crystal. 
This work was a joint studies program of the Institute of Molecular Science (2006), the Use-of-UVSOR Facility Program (BL7B, 2005) and was partially supported by a Grant-in-Aid for Scientific Research (B) (Grant Nos. 18340110, 22340107) from the Ministry of Education, Culture, Sports, Science and Technology of Japan.\\

%
%
%\bibliography{JoergBib}

\end{document}